\documentstyle[12pt,epsf]{article}
\begin{document}
\begin{titlepage}
\rightline{\vbox{\halign{&#\hfil\cr
&UQAM-PHE-97/03\cr
&CUMQ/HEP 96\cr
&\today\cr}}}
\vspace{0.5in}
\begin{center}
{\bf DOUBLY CHARGED HIGGSINO CONTRIBUTION TO THE DECAYS 
$\mu\rightarrow e\gamma$\ AND $\mu\rightarrow 3e$\ 
AND TO THE ANOMALOUS MAGNETIC MOMENT OF
THE MUON $\Delta a_\mu$\ WITHIN THE LEFT-RIGHT SUPERSYMMETRIC MODEL}
\\
\medskip
\vskip0.5in

\normalsize {{\bf G. Couture}$^{\rm a}$,
{\bf M. Frank}$^{\rm b}$\ and {\bf H. K\"{o}nig}$^{\rm a,b}$}
\smallskip
\medskip

{ \sl $^a$
D\'{e}partement de Physique, Universit\'{e} du Qu\'{e}bec \`{a} Montr\'{e}al\\ 
C.P. 8888, Succ. Centre Ville, Montreal, Qu\'{e}bec, Canada, H3C 3P8\\
$^b$Department of Physics, Concordia University, 
1455 De Maisonneuve Blvd. W.\\ Montreal, Quebec, Canada, H3G 1M8 } 
\smallskip
\end{center}
\vskip1.0in

\noindent{\large\bf Abstract}
\smallskip

We present a detailed and complete calculation of the doubly 
charged Higgsino contribution to the decays
 $\mu\rightarrow e\gamma$\ and $\mu\rightarrow 3e$\
and to the anomalous magnetic moment of the muon $\Delta a_\mu$\
within the left-right supersymmetric model. 
We include the mixing 
of the scalar partners of the left and right
handed leptons, and show that it leads to a strong enhancement 
of the decay modes in certain scenarios. 
We find that the contribution of the doubly charged Higgsino
can be close to the known experimental values and are reachable by  
future experiments. 

\end{titlepage}

\baselineskip=20pt

\newpage
\pagenumbering{arabic}
\section{\bf Introduction}
\label{intro}

The quest for a supersymmetric grand unified theory is plagued by lack of direct
 signals which would distinguish such a theory from supersymmetry in general.
Supersymmetry, in particular the Minimal Supersymmetric Model (MSSM)
~\cite{habnil}, can be
probed experimentally through the  high energy production of superpartners.
 However, the MSSM, while filling in some of the theoretical gaps of the 
Standard Model, fails to explain other phenomena such as the weak mixing angle, 
the small mass (or masslessness) of the known neutrinos, the origin of CP 
violation, to quote a few. Extended gauge structures such as grand unified 
theories, introduced to provide an elegant framework for unification of 
forces~\cite{ellis}, would connect 
the standard model with more fundamental structures such as superstrings,
and also resolve the puzzles of the electroweak theory.

Phenomenologically, grand unified theories would either predict relationships 
between otherwise independent parameters of the standard model, 
or new interactions (i.e. 
interactions forbidden or highly suppressed in the standard model).

Among supersymmetric grand unified theories, SO(10)~\cite{barbieri1} and 
SU(5)~\cite{barbieri2} have received significant attention.
 In this aricle we  shall study a model based on the left-right 
symmetric extension of the MSSM, based on $SU(2)_{L}\times SU(2)_{R}\times U(1)_
{B-L}$. Its attraction is that on one hand the Left-Right Supersymmetric 
Model $(LRSUSY)$is an extension of the Minimal 
Supersymmetric Standard Model  based on left-right symmetry and
on the other hand
 it could  be viewed as a low-energy 
realization of certain $SUSY-GUTs$, such as $SO(10)$.

$LRSUSY$ shares some of the attractive  properties of 
the $MSSM$ , like providing a natural solution for the gauge hierarchy
problem. $LRSUSY$ also supresses naturally rapid proton decay,
by disallowing terms in the Lagrangian that explicitly violate either 
baryon or lepton numbers~\cite{frank1}. It gauges the only quantum 
number left ungauged, $B-L$. The $LRSUSY$ model also shares some of 
the attractive 
features of the original left-right symmetric model~\cite{mohapatra},
such as providing a possible explanation for the smallness of the 
neutrino mass and 
for the origin of parity violation. Recently, this model has received a lot
of attention , because it could offer a 
framework for solving both the strong and the weak CP problem~\cite{rasin}.
Furthermore it has the very attractive feature of automatically conserving
R-parity.

So far, there is no 
experimental evidence for the right-handed interactions predicted by the 
$SU(2)_{L}\times SU(2)_{R}\times U(1)_{B-L}$ theory, let alone by supersymmetry.
Yet the foundation of $LRSUSY$ has so many attractive features that the model 
deserves some experimental and theoretical investigation. The next generation 
of linear colliders will provide an excellent opportunity for such a study. 
The theoretical and experimental challenge lies in finding distinctive features
for the left-right supersymmetric model, which allow it to be distinguished 
from both the SUSY version of the Standard Model and from the non-supersymmetric
version of the left-right theory~\cite{frank3}. Lepton-flavor violation decays
 are just the right type of such phenomena.
The  LRSUSY model provides a natural framework for 
large lepton flavor-violating
effects through two mechanisms: on one hand it can give rise to 
a leptonic decay width of the $Z$-boson through both left-handed and 
right-handed scalar lepton mixing~\cite{hamidian} , 
on the other hand it contains 
lepton-flavor-blind higgsinos which couple to leptons only and enhance 
lepton-flavor violation.

In a recent paper ~\cite{cfkp} we analysed the lepton
flavor -violating decay $\mu\rightarrow e\gamma$, 
whose signature can be reliably calculated and, by the 
structure of the model, showed to be greatly enhanced over the MSSM,
and already accessible at the current experimental accuracy.

The contribution of the doubly charged Higgs boson to
lepton-flavor violation decays  
were considered in ~\cite{lepflviol} and quite strong constraints to
its mass and its Yukawa coupling to the leptons were presented.
Here we will study a new source of enhancement coming from the 
existence, in the supersymmetric sector, of a doubly-charged higgsino. Since the
 measurement of $\mu \rightarrow e \gamma$ has a very precise bound
$B(\mu\rightarrow e\gamma) < 4.9\times 10^{-11}$\
and $B(\mu\rightarrow 3e)< 1.0\times 10^{-12}$\
~\cite{particle data},
we are able to obtain values close 
to the experimental bound and restrict some of 
the parameters in the theory.

A similar contribution from the doubly-charged higgsino 
(with different mass parameters) appears in the calculation of another 
accurately measured quantity, 
the anomalous magnetic moment of the muon (AMMM) 
$a_\mu=(g-2)/2$.
Its experimental value is
$a_{\mu}^{\rm exp} = 1 165 922(9)\times 10^{-9}$~\cite{particle data}.
The measured deviation of the AMMM lies
within a range of 
$-2\times 10^{-8}\le \Delta a_\mu^{\rm exp}
\le 2.6\times 10^{-8}$~\cite{baicom} 


Experiment E821, under
way at the Brookhaven National Laboratory (BNL), is designed to improve the
existent data on the AMMM 
by a factor of $10-20$. When completed, 
it would be possible to test deviations from the 1-loop predictions of
the Minimal Supersymmetric Standard Model, believed to arise most likely
from supersymmetric contibutions.
For completeness, we also investigate the effect of the contribution of the 
doubly-charged higgsino to the AMMM. 
We then restrict the parameter space of the 
left-right supersymmetric model by combining these effects.
 
Our paper is organized as follows:
in section~\ref{lrsusy} we give a brief 
description of the model, followed by the 
numerical analysis of the decay $\mu \rightarrow e \gamma $ in 
section~\ref{muega}  
and the AMMM in section~\ref{ammm}. 
Our conclusions 
are reached in section~\ref{concl}. 
In addition, Appendices~\ref{sec-appa}  and ~\ref{sec-appb} will present 
the detailed analytical calculations.

\section{\bf The Left-Right Supersymmetric Model}
\label{lrsusy}

The $LRSUSY$ model, based on $SU(2)_{L}\times SU(2)_{R}\times U(1)_{B-L}$,
has matter 
doublets for both left- and right- handed fermions and the corresponding left- 
and right-handed scalar partners (sleptons and squarks)~\cite{frank3}.
In the gauge sector, 
corresponding to $SU(2)_{L}$ and $SU(2)_{R}$, there are triplet
gauge bosons $(W^{+,-},W^{0})_{L}$, $(W^{+,-},W^{0})_{R}$ and a singlet gauge
boson $V$ corresponding to $U(1)_{B-L}$, together with their superpartners. 
The Higgs sector of this model 
consists of two Higgs bi-doublets, $\Phi_{u}(\frac{1}{2},\frac{1}{2},0)$ and 
$\Phi_{d}(\frac{1}{2},\frac{1}{2},0)$, which are required to give masses to 
both the up and down quarks. In addition, the spontaneous symmetry breaking of
the group 
$SU(2)_{R}\times U(1)_{B-L}$ to the hypercharge symmetry group $U(1)_{Y}$ is 
accomplished by introducing the Higgs triplet fields $\Delta_{L}(1,0,2)$ 
and $\Delta_{R}(0,1,2)$. The choice of the triplets (versus four doublets)
is preferred because with this choice a large Majorana mass can be generated
for the right-handed neutrino and a small one for the left-handed neutrino~
\cite{mohapatra}.
In addition to the triplets $\Delta_{L,R}$, the model must contain two 
additional triplets $\delta_{L}(1,0,-2)$ and $\delta_{R}(0,1,-2)$ , with 
quantum number $B-L= -2$ to insure cancellation of the anomalies that would 
otherwise occur in the fermionic sector. Given their strange quantum numbers,
the $\delta_{L}$ and $\delta_{R}$ do not couple to any of the particles in the 
theory so their contribution is negligible for any phenomenological studies.

As in the standard model, in order to preserve $U(1)_{EM}$ gauge 
invariance,  only the neutral Higgs fields aquire non-zero vacuum 
expectation values $(VEV's)$. These values are:
\begin{eqnarray}
\langle \Delta_{L} \rangle = \left(\begin{array}{cc}
0&0\\v_{L}&0
\end{array}\right),
\langle \Delta_{R} \rangle = \left (\begin{array}{cc}
0&0\\v_{R}&0
\end{array}\right)~\rm{and}~
\langle \Phi \rangle = \left (\begin{array}{cc}
\kappa&0\\0&\kappa' e^{i\omega}
\end{array}\right).
\nonumber
\end{eqnarray}
$\langle \Phi \rangle$ causes the mixing of $W_{L}$ and $W_{R}$ bosons with $CP$-violating 
phase $\omega$. In order to simplify, we will take the $VEV's$ of the Higgs fields as: $\langle \Delta_{L} \rangle = 0$ and 
\begin{eqnarray}
\langle \Delta_{R} \rangle = \left (\begin{array}{cc}
0&0\\v_{R}&0
\end{array}\right),
\langle \Phi_{u} \rangle = \left (\begin{array}{cc}
\kappa_{u}&0\\0&0
\end{array}\right)~\rm{and}~
\langle \Phi_{d} \rangle = \left (\begin{array}{cc}
0&0\\0&\kappa_{d}
\end{array}\right).
\nonumber
\end{eqnarray}
Choosing $v_{L} =\kappa' =0$ satisfies the more loosely required hierarchy
$v_{R}~\gg~max(\kappa,\kappa')~\gg~v_{L}$ and also the required cancellation 
of flavor-changing neutral currents. The Higgs fields aquire non-zero $VEV's$ 
to break both parity and $SU(2)_{R}$.
In the first stage of breaking the right-handed gauge bosons,
 $W_{R}$ and $Z_{R}$
aquire masses proportional to $v_{R}$ and become much heavier than the usual
(left-handed) neutral gauge bosons $W_{L}$ and $Z_{L}$, which pick up masses 
proportional to $\kappa_{u}$ and $\kappa_{d}$ at the second stage of breaking.
~\cite{frank1}

The supersymmetric sector of the model, 
while preserving left-right symmetry, has four singly-charged
charginos ( corresponding to $\tilde\lambda_{L}, 
\tilde\lambda_{R}, \tilde\phi_{u}$, and
$\tilde\phi_{d}$), in addition to $\tilde\Delta_{L}^-$ , 
$\tilde\Delta_{R}^-$ , $\tilde\delta_{L}^-$ and $\tilde\delta_{R}^-$.
The model also has eleven neutralinos, corresponding to $\tilde\lambda_{Z}$, 
$\tilde\lambda_{Z\prime}$,
$\tilde\lambda_{V}$ ,  $\tilde\phi_{1u}^0$ ,$\tilde\phi_{2u}^0$ ,
$\tilde\phi_{1d}^0$ , $\tilde\phi_{2d}^0$, $\tilde\Delta_{L}^0$,
$\tilde\Delta_{R}^0$ $\tilde\delta_{L}^0$, and
$\tilde\delta_{R}^0$. It has been shown that in the scalar sector, 
the left-triplet 
$\Delta_{L}$ couplings can be neglected in phenomenological analyses of 
muon and tau decays~\cite{pilaftsis}. Although $\Delta_{L}$ 
is not necessary for symmetry breaking~\cite{huitu}, 
it is introduced only for preserving left-right symmetry,
we will not neglect the couplings of $\Delta_{L}$ 
in the fermionic sector, which leads to important consequences
as shown later.

The doubly charged $\Delta_{R}^{--}$ is however very important: it carries 
quantum number $B-L $ of 2 and  couples only to leptons, therefore
breaking lepton-quark universality. It and its supersymmetric partner could,
as will be seen in the next 
section, play an important role in flavor-violating leptonic decays.

In the scalar matter sector, the $LRSUSY$ contains two left-handed and two 
right-handed scalar fermions as partners of the ordinary leptons and quarks, 
which themselves
come in left- and right-handed doublets. In general the left- and right-handed  
scalar leptons will mix together. Some of the effects of this mixings, such as 
enhancement of the AMMM, have been discussed 
elsewhere~\cite{frank1}. Only global lepton-family-number violation
would prevent $\tilde{e}$,
$\tilde{\mu}$ and $\tilde{\tau}$ to mix arbitrarily. Permitting this mixing
to occur, we could expect small effects to occur in the non-supersymmetric 
sector, such as radiative muon or tau decays, in addition to other nonstandard 
effects such as massive neutrino oscillations and violation of lepton number 
itself. But, in general, allowing for the mixing, 
we have six charged-scalar lepton states (involving 15 real angles 
and 10 complex phases) and six scalar neutrinos (also involving 15 real 
angles and 10 complex phases).
In order to reduce the (large) number of parameters
we shall assume in what folows that only two generations of
scalar leptons (the lightest)
mix significantly\footnote{However in the appendix~\ref{sec-appa},
we will present general analytical expressions obtained by 
including the mixing of all generations via a supersymmetric
version of the Kobayashi-Maskawa matrix.}. 

The mixings are as follows:
$\tilde\mu_{L,R}$ and $\tilde e_{L,R}$ with angle $\theta_{L,R}$;
$\tilde\nu_{\mu_{L,R}}$ and $\tilde\nu_{e_{L,R}}$ with angle $\alpha_{L,R}$;
so that, for example:

\begin{eqnarray}
\label{leprot1}
\tilde l_{L_{1}}&=& \tilde\mu_{L}cos\theta_{L} + 
\tilde e_{L}sin\theta_{L},\\      
\label{leprot2}
\tilde l_{L_{2}}&=&-\tilde\mu_{L}sin\theta_{L} + 
\tilde e_{L}cos\theta_{L},
\end{eqnarray}
 
and similar for $\tilde l_{R_{1,2}}$ and $\tilde\nu_{L_{1,2}}$ and
$\tilde\nu_{R_{1,2}}$.   

Furthermore we include the mixing of the scalar partners to the
left and right handed leptons; that is, we take $\tilde e_L=
\cos\theta_{\tilde e} \tilde e_1-\sin\theta_{\tilde e}\tilde e_2$\ 
and $\tilde e_R=
\sin\theta_{\tilde e}\tilde e_1+\cos\theta_{\tilde e} \tilde e_2$, 
and similiar for the
other generations, where we express
the mixings by the matrix $\tilde K^a_{mn}$\
in Appendix~\ref{sec-appa}.

The physical mass eigenstates are then given by
eq.(\ref{leprot1}-\ref{leprot2}).

Next we consider the implications of the above-mixing in the $LRSUSY$ in
lepton-flavor violating decay $\mu \rightarrow e \gamma$.

\section{\bf The Decay $\mu \rightarrow e \gamma$ }
\label{muega}

\begin{figure}[hbtp]
\begin{center}
\mbox{\epsfxsize=144mm\epsffile{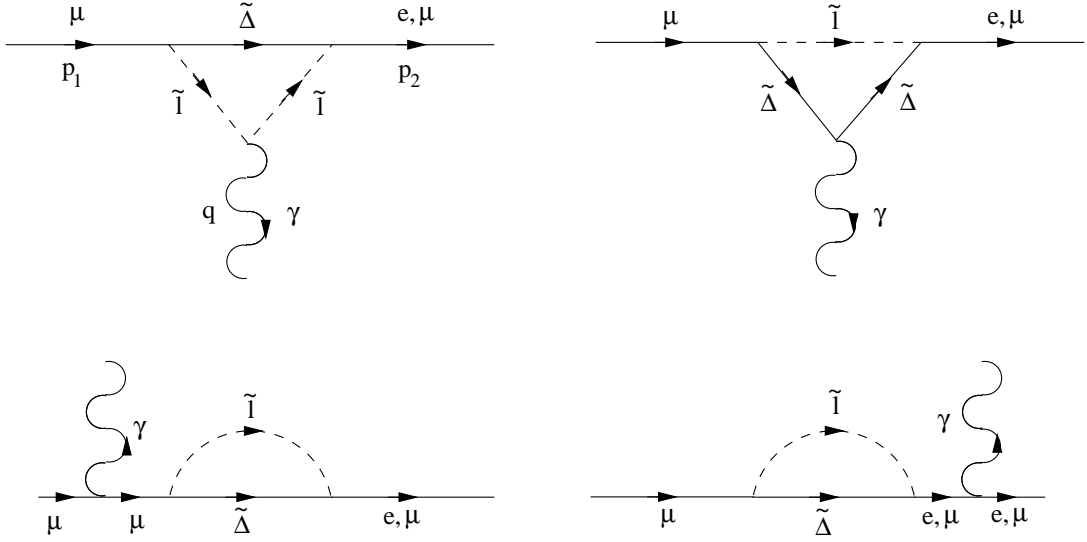}}
\end{center}
\caption{
The penguin diagrams with scalar leptons and
doubly charged Higgsinos within the loop. In the left-right model
without SUSY the same diagrams occur with $\tilde \Delta\leftrightarrow l$\
and $\tilde l\leftrightarrow \Delta_{L,R}$} 
\label{muegf}
\end{figure}

To obtain the contributions of the doubly charged higgsino to 
the decay rate of $\mu\rightarrow e\gamma$\ we have to
calculate the diagrams of fig.(\ref{muegf}).

The detailed calculations are shown in the Appenix~\ref{sec-appa}.
The matrix element given in eq.(\ref{muegeq}) leads to the
following result: 

\begin{eqnarray}
\label{muegfin}
iM&=& +\frac{eh_{LR}^2}{(4\pi)^2}\frac{m_\mu}
{m_{\tilde\Delta}^2}
\overline u_{p_2}i\sigma_{\mu\nu}q^\nu\Bigl\lbrace 
c_{\theta_L} s_{\theta_L} (B^L_{\tilde e}-B^L_{\tilde\mu})P_R+
c_{\theta_R} s_{\theta_R} (B^R_{\tilde e}-B^R_{\tilde\mu})P_L
\Bigr\rbrace u_{p_1}\epsilon^{\ast\mu}_q\\
B^L_{\tilde l}&=&c_{\theta_{\tilde l}}^2 B_{\tilde l_1}+
s_{\theta_{\tilde l}}^2 B_{\tilde l_2}+\frac{m_{\tilde\Delta}}
{m_\mu}
c_{\theta_{\tilde l}}s_{\theta_{\tilde l}}
(\frac{c_{\theta_R}}{c_{\theta_L}}\tilde B_{\tilde l_1}
-\frac{s_{\theta_R}}{s_{\theta_L}}\tilde B_{\tilde l_2})\nonumber\\
B^R_{\tilde l}&=&s_{\theta_{\tilde l}}^2 B_{\tilde l_1}+
c_{\theta_{\tilde l}}^2 B_{\tilde l_2}+\frac{m_{\tilde\Delta}}
{m_\mu}
c_{\theta_{\tilde l}}s_{\theta_{\tilde l}}
(\frac{c_{\theta_L}}{c_{\theta_R}}\tilde B_{\tilde l_1}
-\frac{s_{\theta_L}}{s_{\theta_R}}\tilde B_{\tilde l_2})\nonumber\\
B_{\tilde l_m}&=& e_{\tilde\Delta}\lbrack 
- B_G(x_{\tilde l_m}^{\tilde \Delta})
+B_F(x_{\tilde l_m}^{\tilde \Delta})\rbrack
+e_{\tilde l} B_F(x_{\tilde l_m}^{\tilde \Delta})
\nonumber\\
\tilde B_{\tilde l_m}&=&e_{\tilde\Delta}\tilde B_G(x^{\tilde\Delta}_
{\tilde l_m})+e_\mu\tilde B_F(x^{\tilde\Delta}_{\tilde l_m})
\nonumber
\end{eqnarray}

In order to not to have too many parameters, 
we assume that $h_{LR}$\ of different generations are all of the
same order. We will however keep the difference of the couplings within 
the first and second generation in the 
mixing angles $c_{\theta_{L,R}}$,
eq.(\ref{leprot1}-\ref{leprot2}).
$c_{\theta_{\tilde l}}$\ is the mixing angle of the scalar
partners of the left and right handed leptons.  
Note that if we would include the mixing of all 3 generations
the expression is far more complicated, in particular for the
Higgsino mass term. 
Since $m_{\tilde\Delta}\gg m_l$\ 
the mixing of the scalar leptons can not be neglected. 
If we exclude the contribution of the left handed
Higgsino then $B^L_{\tilde l}$\ and, as can be easily seen  
from eq.(\ref{muegeq}) (all $\tilde K_{m1}^a$\ set to be 0), 
the term proportional to the Higgsino mass are absent 
\footnote{
The calculation of the diagrams 
with the doubly charged left and right Higgs bosons
and leptons within the loop, which occur in the left-right model
without SUSY are very similiar ($e_{\tilde\Delta}\leftrightarrow e_{l^C}=+1$,
$e_{\tilde l}\leftrightarrow e_{\Delta_{L,R}}$\ and
$x_{\tilde l_m}^{\tilde \Delta}\leftrightarrow x^l_{\Delta_{L,R}}=
m_{\Delta_{L,R}}^2/m_l^2$). Limits of the couplings and the masses
of the doubly charged Higgs boson have been given elsewhere and will
not presented here ~\cite{lepflviol}.}.

A matrix element of the form $iM=e\overline u_{p_2}i\sigma_{\mu\nu}
q^\nu (a_L P_L+a_R P_R) u_{p_1}\epsilon^{\ast\mu}_q$\ leads to
a decay rate of 
$\Gamma(\mu\rightarrow e\gamma)=m_\mu^3(a_L^2+a_R^2)\alpha/4$\
and $\Gamma(\mu\rightarrow 3e)=\frac{1}{3}\frac{\alpha^2}{(4\pi)}
(a_L^2+a_R^2)m_\mu^3\lbrack\log\frac{m_\mu^2}{m_e^2}-\frac{11}{4}\rbrack$.
With $\Gamma(\mu\rightarrow\nu_\mu e^-\overline\nu_e)=
\frac{G_F^2}{192\pi^3}m_\mu^5=\frac{\alpha^2}{384\pi s^4_W}
\left(\frac{m_\mu}{m_W}\right)^4m_\mu$, where $s_W$\ is the Weinberg
angle $\sin\theta_W$, we obtain the following branching ratio
for eq.(\ref{muegfin}):

\begin{equation}
\label{bmueg}
B(\mu\rightarrow e\gamma)=\frac{3s_W^4}{8\pi^3}\frac{h^4_{LR}}
{\alpha}\left(\frac{m_W}{m_{\tilde\Delta}}\right)^4
\Bigl\lbrack
c^2_{\theta_L} s^2_{\theta_L}(B^L_{\tilde e}-B^L_{\tilde\mu})^2
+c^2_{\theta_R} s^2_{\theta_R}(B^R_{\tilde e}-B^R_{\tilde\mu})^2
\Bigr\rbrack
\end{equation}

Simliar for $B(\mu\rightarrow 3e)$. It is 
$\frac{\Gamma(\mu\rightarrow 3e)}{\Gamma(\mu\rightarrow e\gamma)}=
\frac{\alpha}{3\pi}\lbrack \log\frac{m_\mu^2}{m_e^2}-\frac{11}{4}\rbrack$.

If we neglect the mixing of the scalar partners of the left
and right handed leptons, that is, setting
$\sin\theta_{\tilde l}=0$\ eq.(\ref{bmueg}) leads to:

\begin{equation}
\label{bmueg0}
B(\mu\rightarrow e\gamma)=\frac{3s_W^4}{8\pi^3}\frac{h^4_{LR}}
{\alpha}\left(\frac{m_W}{m_{\tilde\Delta}}\right)^4
\Bigl\lbrack
c^2_{\theta_L} s^2_{\theta_L}(B_{\tilde e_1}-B_{\tilde\mu_1})^2
+c^2_{\theta_R} s^2_{\theta_R}(B_{\tilde e_2}-B_{\tilde\mu_2})^2
\Bigr\rbrack
\end{equation}

The finite result depends strongly on
the mass differences of the first and second generation.
If we take as, required by left-right symmetry 
$\theta_L\equiv\theta_R\equiv\theta$\ and 
$B_{\tilde l_1}=B_{\tilde l_2}$\ with $\Delta m_{\tilde l}^2=
\Delta m_l^2$\ for different generations 
and $m_{\tilde\Delta}\sim m_{\tilde l}$, a first
estimate is given by:

\begin{equation}
\label{bmuegest}
B(\mu\rightarrow e\gamma)=\frac{3s_W^4}{4\pi^3}\frac{h^4_{LR}}
{\alpha}\left(\frac{m_W}{m_{\tilde\Delta}}\right)^4
c^2_{\theta}s^2_{\theta}
\left(\frac{7}{120}\right)^2
\left(\frac{m_\mu}{m_{\tilde\Delta}}\right)^4\\
\end{equation}

Eq.(\ref{bmuegest}) leads to $B(\mu\rightarrow e\gamma)\approx
2.97\times 10^{-16}h^4_{LR}c^2_{\theta}s^2_{\theta}$\
for $m_{\tilde\Delta}=100$\ GeV and thus far smaller than 
the experimental limit.

However the situation is changed when the mixing of the 
scalar partners of the right and left handed leptons
is included. Since $m_l\ll m_{\tilde l}$\ the mixing
angle is given by \hfill\break
$\sin\theta_{\tilde l}\sim m_l(A_l-\mu\tan\beta)/m_{\tilde l}^2
\sim m_l/m_{\tilde l}$\ with $(A_l-\mu\tan\beta)\sim m_{\tilde l}$,
where $A_l$\ is the trilinear scalar interaction, $\mu$\ the
mixing mass terms of the Higgs bosons and $\tan\beta=v_2/v_1$\
the ratio of their vacuum expectation values (vev's). This
leads to $B^L_{\tilde l}\approx B_{\tilde l_1}+
\frac{m_{\tilde\Delta}}{m_{\tilde l}}
\frac{m_l}{m_\mu}(\tilde B_{\tilde l_1}-
\tilde B_{\tilde l_2})$\ where we assumed that after left-right
symmetry breaking the relations $c_{\theta_L}/c_{\theta_R}\sim
s_{\theta_L}/s_{\theta_R}\sim 1$\ still hold. 

If the left-right symmetry is conserved we have
$\tilde B_{\tilde l_1}=\tilde B_{\tilde l_2}$\
and eq.(\ref{bmueg0}) is recovered. However after
breaking, the mass difference between the scalar partners
of the left handed and right handed leptons is much larger
than the mass difference between the generations. We expect  
$\frac{m_{\tilde l_1}^2-m_{\tilde l_2}^2}{m_{\tilde l}^2}\sim
10^{-2}-10^{-1}$\ ~\cite{rasin,posp}. 
That is $B_{\tilde e}^{L,R}-B_{\tilde u}^{L,R}\sim
\frac{m_{\tilde\Delta}}{m_{\tilde l}}
(\tilde B_{\mu_2}-\tilde B_{\mu_1})\sim \frac{5}{12}(10^{-2}-10^{-1})$\
with $m_{\tilde\Delta}\sim m_{\tilde l}$. With these values
eq.(\ref{bmueg}) leads to $B(\mu\rightarrow e\gamma)\approx
1.25\times (10^{-6}-10^{-4})h^4_{LR}c^2_{\theta}s^2_{\theta}$\
for $m_{\tilde\Delta}=100$\ GeV and thus far above the
known experimental limit of $4.9\times 10^{-11}$, leading to
the constraint $h_{LR}^2<(10^{-9}-10^{-8})m_{\tilde\Delta}^2 $\ 
(GeV) when maximal
electron-muon mixing is assumed.

\section{\bf The Anomalous Magnetic Moment of the Muon}
\label{ammm}

The calculations to the AMMM are similiar as to the $\mu\rightarrow
e\gamma$\ decay. From eq.(\ref{anmmeq}) we obtain the following
result:

\begin{equation}
\label{anmmfin}
\Delta a_\mu=-\frac{2h_{LR}^2}{(4\pi)^2}\left (\frac{m_\mu}
{m_{\tilde\Delta}}\right )^2\Bigl\lbrace 
B^L_{\tilde\mu}+B^R_{\tilde\mu}\Bigr\rbrace
\end{equation} 

$B^L_{\tilde\mu}$\ and $B^R_{\tilde\mu}$\ as given in
eq.(\ref{muegfin}). Again $B^L_{\tilde\mu}$\ is absent when
we exclude the contribution of the left handed higgsino.

Eq.(\ref{anmmfin}) leads to

\begin{equation}
\label{anmmest}
\Delta a_\mu=-\frac{2h_{LR}^2}{(4\pi)^2}\left(\frac{m_\mu}
{m_{\tilde\Delta}}\right)^2(B_{\tilde\mu_1}+B_{\tilde\mu_2}) 
\end{equation}

Eq.(\ref{anmmest}) leads to 
$\Delta a_\mu\approx -3.49\times 10^{-9}
h^2_{LR}$\
for $m_{\tilde\Delta}\sim m_{\tilde\mu}=100$\ GeV, thus already
very close to the experimental lower limit of
$\Delta a_\mu^{\rm exp}$, leading to the constraint
$h_{LR}^2<5\cdot 10^{-4}m_{\tilde\Delta}^2$\ (GeV).


\section{\bf Conclusions}
\label{concl}
We presented the results of the contribution of the doubly
charged Higgsino to the decays $\mu\rightarrow e\gamma$\
and $\mu\rightarrow 3e$\ and to the anomalous magnetic moment
of the muon $\Delta a_\mu$. It was shown that when mixing
of the scalar partners of the left and right handed 
leptons are neglected the contribution to the decay modes is 
far below the experimental limits. However 
for the anomalous magnetic moment it is within the reach
of the future BNL experiments.
When mixing of the scalar leptons is included
we obtain a constrain for the lepton-scalar lepton-doubly
charged Higgsino coupling $h_{LR}$\ in the order of less
than $10^{-4}m_{\tilde\Delta}$\ (GeV)
 if maximal electron-muon mixing is assumed
(not to be confused with the upper mixing of the scalar leptons).
The effect of this mixing on $\Delta a_\mu$\ is negligible.

\section{\bf Acknowledgements}

We want to thank M. Pospelov
for fruitful discussions.
This work was funded by NSERC of Canada and les Fonds FCAR du Qu\'ebec.

\appendix
\section{Appendix: The matrix elements }
\label{sec-appa}
 
In this Appendix we present the final results of the calculations of
the Feynman diagrams as shown in Fig.1 with 
flavor non diagonal couplings of the doubly charged Higgsino to the
left and right handed leptons and their scalar partners.
We define the following matrix:

\begin{equation}
iM=+{{eh_{LR}^2}\over{(4\pi)^2}}(M_S+M_F+M_{SE})
\label{defeq}
\end{equation}

where $M_S$\ is the matrix element for the two scalar leptons
in the loop, $M_F$\ the one for the two doubly charged Higgsinos
in the loop and $M_{SE}$\ the one for the self energy diagrams.
The calculations of the diagrams are very similiar to those
presented in the Appendix~\ref{sec-appa} of ~\cite{hk1}; where we refer the
interested reader to for more details. Here we only present
the finite results for each diagram.  

\begin{eqnarray}
\label{fermeq}
iM_F&=&\int_0^1 d\alpha_1 e_{\tilde\Delta}\biggl\lbrace
\overline u_{p_2}(q^2\gamma_\mu-\rlap/q q_\mu)(T_{11}^{aL}P_L+
T_{22}^{aR}P_R)u_{p_1}{1\over 6}(2-3\alpha_1+\alpha^3_1)
\\
& &+\overline u_{p_2}i\sigma_{\mu\nu}q^\nu\Bigl\lbrace
m_{\tilde\Delta}(1-\alpha_1)(T_{12}^{aL}P_L+T_{12}^{aR}P_R)
\nonumber \\
& &-{1\over 2}\alpha_1(1-\alpha_1)^2\bigl\lbrack m_{e,\mu}
(T_{11}^{aL}P_L+T_{22}^{aR}P_R)+m_\mu
(T_{11}^{aL}P_R+T_{22}^{aR}P_L)\bigr\rbrack\Bigr\rbrace u_{p_1}
\biggr\rbrace 
{{\epsilon^{\ast\mu}_q}\over{D_{SE}^{\tilde\Delta\tilde l}}}
\nonumber\\
\label{scaeq}
iM_S&=&\int_0^1 d\alpha_1 e_{\tilde l}\biggl\lbrace
\overline u_{p_2}(q^2\gamma_\mu-\rlap/q q_\mu)(T_{11}^{aL}P_L+
T_{22}^{aR}P_R)u_{p_1}{1\over 6}\alpha_1^3
\\
& &+\overline u_{p_2}i\sigma_{\mu\nu}q^\nu\Bigl\lbrack
m_{e,\mu}(T_{11}^{aL}P_L+T_{22}^{aR}P_R)
+m_\mu(T_{11}^{aL}P_R+T_{22}^{aR}P_L)\Bigr\rbrack
u_{p_1}{1\over 2}\alpha_1^2(1-\alpha_1)\biggr\rbrace
{{\epsilon^{\ast\mu}_q}\over{D_{SE}^{\tilde\Delta\tilde l}}}
\nonumber\\
\label{selfeq}
iM_{SE}&=&\int_0^1 d\alpha_1 \Bigl\lbrack-e_\mu\Bigl\lbrace
u_{p_2}i\sigma_{\mu\nu}q^\nu
\bigl\lbrack -\alpha_1(1-\alpha_1)m_{\tilde\Delta}
(T_{12}^{aL}P_L+T_{12}^{aR}P_R)\bigr\rbrack u_{p_1}\Bigr\rbrace
\Bigr\rbrack 
{{\epsilon^{\ast\mu}_q}\over{D_{SE}^{\tilde\Delta\tilde l}}}
\\
D_{SE}^{\tilde\Delta\tilde l}&=&m_{\tilde\Delta}^2-(m_{\tilde\Delta}^2
-m^2_{{\tilde l}_
{a_m}})\alpha_1
\nonumber \\
T_{11}^{aL}&:=&K^{\ast L}_{ae,\mu}K^L_{a\mu}\tilde K_{m1}^{a2}
\quad
T_{22}^{aR}:=K^{\ast R}_{ae,\mu}K^R_{a\mu}\tilde K_{m2}^{a2}
\nonumber \\
T_{12}^{aL}&:=&K^{\ast R}_{ae,\mu}K^L_{a\mu}\tilde K_{m1}^a
\tilde K_{m2}^a
\quad
T_{12}^{aR}:=K^{\ast L}_{ae,\mu}K^R_{a\mu}\tilde K_{m1}^a
\tilde K_{m2}^a\nonumber
\end{eqnarray}

where $e_{\tilde\Delta}=-2$, $e_{\tilde l}=+1$\ and $e_\mu=-1$. 
We have to sum over the generation indices $a=1-3$\ and over
the eigenstates of the scalar partners of the left and right
handed leptons $m=1,2$.
The divergencies
cancel after the summation of all diagrams 
$e_{\tilde\Delta}+e_{\tilde l}-e_\mu\equiv 0$. Due to this equation many more
terms like terms proportional to 
$m_{\tilde\Delta}\alpha_1(1-\alpha_1)
(p_1+p_2)^\mu$\ drop out after summation and are not explicitely
written down in eqs.(\ref{fermeq}-\ref{selfeq}). 
Furthermore we made use of
eqs.(A.5)-(A.8) in ~\cite{hk1} and eqs.(A.6)+(A.7) in ~\cite{hk2}.
After summation of eqs.(\ref{fermeq}-\ref{selfeq}) 
the final matrix element is given by:

\begin{eqnarray}
iM&=&+{{eh_{LR}^2}\over{(4\pi)^2}}{1\over{m_{\tilde\Delta}^2}}
\biggl\lbrace 
\overline u_{p_2}(q^2\gamma_\mu-\rlap/q q_\mu)(T_{11}^{aL}P_L+
T_{22}^{aR}P_R)u_{p_1}
\\
& &\Bigl\lbrace e_{\tilde\Delta}\lbrack
A_G(x_{\tilde l_{a_m}}^{\tilde\Delta})+
A_F(x_{\tilde l_{a_m}}^{\tilde\Delta })\rbrack
+e_{\tilde l} A_F(x_{\tilde l_{a_m}}^{\tilde\Delta })
\Bigr\rbrace\nonumber \\
& &+\overline u_{p_2}i\sigma_{\mu\nu}q^\nu\Bigl\lbrace
m_{\tilde\Delta}(T_{12}^{aL}P_L+T_{12}^{aR}P_R)
\lbrack e_{\tilde\Delta}
\tilde B_G(x_{\tilde l_{a_m}}^{\tilde\Delta })
+e_\mu\tilde B_F(x_{\tilde l_{a_m}}^{\tilde\Delta})\rbrack
\nonumber\\
& &\bigl\lbrack m_{e,\mu}
(T_{11}^{aL}P_L+T_{22}^{aR}P_R)+m_\mu
(T_{11}^{aL}P_R+T_{22}^{aR}P_L)\bigr\rbrack
\nonumber \\
& &\bigl\lbrack e_{\tilde\Delta}\lbrack 
- B_G(x_{\tilde l_{a_m}}^{\tilde \Delta})
+B_F(x_{\tilde l_{a_m}}^{\tilde \Delta})\rbrack
+e_{\tilde l} B_F(x_{\tilde l_{a_m}}^{\tilde \Delta})\bigr\rbrack
\Bigr\rbrace u_{p_1}\biggr\rbrace\epsilon^{\ast\mu}_q
\nonumber
\label{finres}
\end{eqnarray}

The final functions after Feynman integration are shown in Appendix 
\ref{sec-appb}. Since we are not interested in CP violation we can
skip the $\ast$\ in $T_{kl}^{aL,R}$
\footnote{We could also have skipped the $L$\ and $R$\ indices
(that is $\theta_L=\theta_R$) since the difference between the
left and right diagonalizing matrices lies in a relative minus sign
of the SUSY phase, see eq.(4) in~\cite{gegriray}. This would simplify 
the results strongly giving us an overall factor of $K_{ae,\mu}K_{a\mu}$\
and $T_{12}^{aL}\equiv T_{12}^{aR}$. For the moment however we will
keep $\theta_L$\ different from $\theta_R$.}.
Furthermore for
a photon on mass shell we have $q^2=0$\ and due to gauge invariance
${\epsilon^{\ast\mu}_q}q_\mu=0$. 

Neglecting the electron mass
the matrix element for the decay $\mu\rightarrow e\gamma$\ is then
given by:

\begin{eqnarray}
\label{muegeq}
iM&=&+{{eh_{LR}^2}\over{(4\pi)^2}}{1\over{m_{\tilde\Delta}^2}}
\overline u_{p_2}i\sigma_{\mu\nu}q^\nu\Bigl\lbrace
m_\mu\bigl\lbrack K^L_{a1}K^L_{a2}\tilde K_{m1}^{a2}P_R+
K^R_{a1}K^R_{a2}\tilde K_{m2}^{a2}P_L\bigr\rbrack\\& &
\times\Bigl\lbrace e_{\tilde\Delta}\lbrack 
- B_G(x_{\tilde l_{a_m}}^{\tilde \Delta})
+B_F(x_{\tilde l_{a_m}}^{\tilde \Delta})\rbrack
+e_{\tilde l} B_F(x_{\tilde l_{a_m}}^{\tilde \Delta})\Bigr\rbrace
\nonumber\\
& &+m_{\tilde\Delta} 
\tilde K_{m1}^a\tilde K_{m2}^a
(K^R_{a1}K^L_{a2}P_R+K^L_{a1}K^R_{a2}P_L)
\lbrack e_{\tilde\Delta}
\tilde B_G(x_{\tilde l_{a_m}}^{\tilde\Delta })
+e_\mu\tilde B_F(x_{\tilde l_{a_m}}^{\tilde\Delta})\rbrack
\Bigr\rbrace u_{p_1}\epsilon^{\ast\mu}_q
\nonumber
\end{eqnarray}

After summation over the mass eigenstates of the scalar leptons and 
all the generations making use of the unitarity
of the leptonic KM matrix
eq.(\ref{muegeq}) leads to the  branching ratio 
$B(\mu\rightarrow e\gamma)$\
 as given in eq.(\ref{muegfin}).

The ANMM can be extracted from 
$V_\mu={{ie}\over{2m_\mu}}F(q^2)\overline u_{p_2}i\sigma_{\mu\nu}
q^\nu u_{p_1}$\ at $q^2=0$. 
We obtain the result: 

\begin{eqnarray}
\label{anmmeq}
F(q^2)\vert_{q^2=0}=\Delta a_\mu&=&-{{2h_{LR}^2}\over{(4\pi)^2}}
\left ({{m_\mu}\over{m_{\tilde\Delta}}}\right )^2
\biggl\lbrace
\Bigl\lbrack\vert K^L_{a2}\vert^2\tilde K_{m1}^{a2}+
\vert K^R_{a2}\vert^2\tilde K_{m2}^{a2}\Bigr\rbrack
\\& &
\times\Bigl\lbrack e_{\tilde\Delta}\lbrack 
- B_G(x_{\tilde l_{a_m}}^{\tilde \Delta})
+B_F(x_{\tilde l_{a_m}}^{\tilde \Delta})\rbrack
+e_{\tilde l} B_F(x_{\tilde l_{a_m}}^{\tilde \Delta})\Bigr\rbrack\quad
\nonumber\\
& &+{{m_{\tilde\Delta}}\over{m_\mu}}
K^R_{a2}K^L_{a2}
\tilde K_{m1}^a\tilde K_{m2}^a
\lbrack e_{\tilde\Delta}
\tilde B_G(x_{\tilde l_{a_m}}^{\tilde\Delta })
+e_\mu\tilde B_F(x_{\tilde l_{a_m}}^{\tilde\Delta})\rbrack
\biggr\rbrace\nonumber
\end{eqnarray}

As above after summation over all mass eigenstates 
and generations of the leptons 
eq.({\ref{anmmeq}) leads to the result as given in eq.(\ref{anmmfin}). 

\section{Appendix: The final functions }
\label{sec-appb}

\begin{eqnarray}
A_G(x_{\tilde l_{a_m}}^{\tilde\Delta})&=&{1\over 6}\int\limits_0^1
d\alpha_1\bigl\lbrack 2-3\alpha_1^2-3\alpha_1(1-\alpha_1)
\bigr\rbrack/ D_{\tilde\Delta  \tilde l_{a_m}}
\\
&=&{1\over{(1-x_{\tilde l_{a_m}}^{\tilde\Delta })^2}}{1\over 2}
\bigl\lbrace 1-x_{\tilde l_{a_m}}^{\tilde\Delta }+{1\over 3}(1+2
x_{\tilde l_{a_m}}^{\tilde\Delta})
\log(x_{\tilde l_{a_m}}^{\tilde \Delta})\bigr\rbrace
\nonumber\\
A_F(x_{\tilde l_{a_m}}^{\tilde\Delta })&=&{1\over 6}\int\limits_0^1
d\alpha_1\alpha_1^3/D_{\tilde\Delta  \tilde l_{a_m}}
\\
&=&{1\over{(1-x_{\tilde l_{a_m}}^{\tilde \Delta})^4}}{1\over 36}
\bigl\lbrace -11+18 x_{\tilde l_{a_m}}^{\tilde \Delta}
-9x_{\tilde l_{a_m}}^{\tilde \Delta\ 2}
+2x_{\tilde l_{a_m}}^{\tilde \Delta\ 3}-
6\log(x_{\tilde l_{a_m}}^{\tilde \Delta})\bigr\rbrace
\nonumber\\
B_G(x_{\tilde l_{a_m}}^{\tilde \Delta})&=&{1\over 2}\int\limits_0^1
d\alpha_1\alpha_1(1-\alpha_1)/D_{\tilde \Delta \tilde l_{a_m}}
\\
&=&{1\over{(1-x_{\tilde l_{a_m}}^{\tilde\Delta })^3}}{1\over 4}
\bigl\lbrace 1-x_{\tilde l_{a_m}}^{\tilde \Delta\ 2}+
2x_{\tilde l_{a_m}}^{\tilde \Delta}\log(
x_{\tilde l_{a_m}}^{\tilde \Delta})\bigr\rbrace
\nonumber\\
B_F(x_{\tilde l_{a_m}}^{\tilde \Delta})&=&{1\over 2}\int\limits_0^1
d\alpha_1\alpha_1^2(1-\alpha_1)/D_{\tilde\Delta \tilde l_{a_m}}\\
&=&{1\over{(1-x_{\tilde l_{a_m}}^{\tilde\Delta})^4}}{1\over 12}\bigl\lbrace
2+3x_{\tilde l_{a_m}}^{\tilde \Delta}-
6x_{\tilde l_{a_m}}^{\tilde\Delta\ 2}+x_{\tilde l_{a_m}}^{\tilde\Delta\ 3}
+6x_{\tilde l_{a_m}}^{\tilde\Delta}
\log(x_{\tilde l_{a_m}}^{\tilde\Delta})\bigr\rbrace
\nonumber\\
\tilde B_G(x_{\tilde l_{a_m}}^{\tilde\Delta })&=&\int\limits_0^1d\alpha_1
(1-\alpha_1)/D_{\tilde\Delta \tilde l_{a_m}}\\
&=&{1\over{(1-x_{\tilde l_{a_m}}^{\tilde \Delta})^2}}
\bigl\lbrace 1-x_{\tilde l_{a_m}}^{\tilde \Delta}
+x_{\tilde l_{a_m}}^{\tilde \Delta}
\log(x_{\tilde l_{a_m}}^{\tilde\Delta })\bigr\rbrace
\nonumber\\
\tilde B_F(x_{\tilde l_{a_m}}^{\tilde\Delta})&=
&2 B_G(x_{\tilde l_{a_m}}^{\tilde \Delta})\\
D_{\tilde\Delta \tilde l_{a_m}}&=&1-(1-x_{\tilde l_{a_m}}^{\tilde \Delta})
\alpha_1\nonumber\\
x_{\tilde l_{a_m}}^{\tilde \Delta}&=&{{m_{\tilde l_{a_m}}^2}\over
{m_{\tilde\Delta}^2}}
\nonumber
\label{fineq}
\end{eqnarray}

We have the following relations between the eqs(B.3-B.6) and the eqs.(4.10),
(4.11),(4.17) and (4.21) of ~\cite{frank1}.
$\frac{m_\mu^2}{m_{\tilde\Delta}^2}B_F=G'/6$, $\frac{m_\mu}{m_{\tilde\Delta}}
\tilde B_F=G/2$, $\frac{m_\mu^2}{m_{\tilde\Delta}^2}(B_G-B_F)=F'/12$\
and $\frac{m_\mu}{m_{\tilde\Delta}}(\tilde B_G-\tilde B_F)=F/2$. 
Furthermore we have
$\frac{m_\mu}{m_{\tilde\Delta}}G=(2G'+F')/3$.

\end{document}